\documentclass[lettersize,journal]{IEEEtran}
\usepackage{amsmath,amsfonts}
\usepackage{algorithmic}
\usepackage{algorithm}
\usepackage{array}
\usepackage[caption=false,font=normalsize,labelfont=sf,textfont=sf]{subfig}
\usepackage{textcomp}
\usepackage{stfloats}
\usepackage{url}
\usepackage[table]{xcolor}
\usepackage{xcolor}
\usepackage{verbatim}
\usepackage{siunitx}
\usepackage{graphicx}
\usepackage{cite}
\usepackage{booktabs}
\usepackage{adjustbox}
\usepackage{amssymb}
\usepackage{placeins}
\usepackage{newfloat}
\usepackage{amsmath}
\usepackage{listings}
\usepackage{gensymb}
\begin{document}

\title{FlexHDR: Modelling Alignment and Exposure Uncertainties for Flexible HDR Imaging}

\author{Sibi Catley-Chandar, Thomas Tanay, Lucas Vandroux, Ale\v{s} Leonardis, Gregory Slabaugh, Eduardo P\'{e}rez-Pellitero
\thanks{Sibi Catley-Chandar, Thomas Tanay, Lucas Vandroux, Ale\v{s} Leonardis, and Eduardo P\'{e}rez-Pellitero are with Huawei Noah's Ark Lab.}
\thanks{Sibi Catley-Chandar and Gregory Slabaugh are with Queen Mary University of London.}}

\makeatletter
\def\ps@IEEEtitlepagestyle{%
  \def\@oddfoot{\mycopyrightnotice}%
  \def\@oddhead{\hbox{}\@IEEEheaderstyle\leftmark\hfil\thepage}\relax
  \def\@evenhead{\@IEEEheaderstyle\thepage\hfil\leftmark\hbox{}}\relax
  \def\@evenfoot{}%
}
\def\mycopyrightnotice{%
  \begin{minipage}{\textwidth}
  \centering \scriptsize
  Copyright~\copyright~2022 IEEE.  Personal use of this material is permitted.  Permission from IEEE must be obtained for all other uses, in any current or future media, including reprinting/republishing this material for advertising or promotional purposes, creating new collective works, for resale or redistribution to servers or lists, or reuse of any copyrighted component of this work in other works.
  \end{minipage}
}
\makeatother

\maketitle


\begin{abstract}
High dynamic range (HDR) imaging is of fundamental importance in modern digital photography pipelines and used to produce a high-quality photograph with well exposed regions despite varying illumination across the image. This is typically achieved by merging multiple low dynamic range (LDR) images taken at different exposures. However, over-exposed regions and misalignment errors due to poorly compensated motion result in artefacts such as ghosting. In this paper, we present a new HDR imaging technique that specifically models alignment and exposure uncertainties to produce high quality HDR results.
We introduce a strategy that learns to jointly align and assess the alignment and exposure reliability using an HDR-aware, uncertainty-driven attention map that robustly merges the frames into a single high quality HDR image.  Further, we introduce a progressive, multi-stage image fusion approach that can flexibly merge any number of LDR images in a permutation-invariant manner. Experimental results show our method can produce better quality HDR images with up to 1.1dB PSNR improvement to the state-of-the-art, and subjective improvements in terms of better detail, colours, and fewer artefacts. 
\end{abstract}




\section{Introduction}
\IEEEPARstart{D}{espite} recent advances in imaging technology, capturing scenes with wide dynamic range still poses several challenges. Current camera sensors suffer from limited or Low Dynamic Range (LDR) due to inherent hardware limitations. 
The maximum dynamic range a camera can capture is closely related to (a) the sensor's photosite \textit{full well electron capacity} or saturation point, and (b) the black point, which is generally constrained by the uncertainty in the reading due to the dominant presence of noise.

Different solutions have been proposed to overcome these limitations. The principle behind most of them relies on capturing observations of the same scene with different exposure values. This enables a richer coverage of the scene's original dynamic range, but also requires a mechanism to align and unify the different captured observations~\cite{Debevec97}. Some approaches 
make use of multi-sensor or multi-camera configurations, e.g.~Tocci et al.~\cite{Tocci11}, McGuire et al.~\cite{McGuire07}, Froehlich et al.~\cite{Froehlich14}, where a beam splitter enables the light to be captured by multiple sensors.
However, such setups are normally expensive, fragile, with bulky and cumbersome rigs, and they may suffer from double contours, light flares, or polarization artefacts~\cite{Froehlich14}.

\begin{figure}[t]
\centering
\includegraphics[width=1.0\columnwidth]{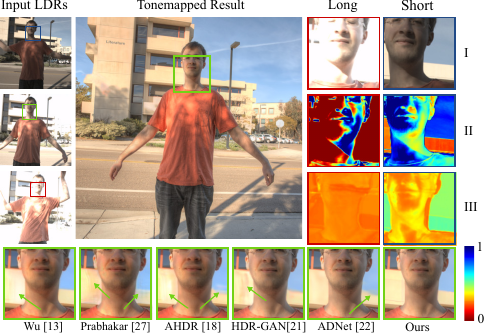}
\caption{
Our method, intermediate results, and comparisons.
LDR input images are shown on the left and our tone mapped result is in the centre. The visualisations on the right are: \textbf{I} Input Image, \textbf{II} Exposure Map, \textbf{III} Attention Map. 
The bottom row compares to several state-of-the-art methods. Our uncertainty modelling more effectively handles regions of overexposure and motion between input frames. 
}
\label{fig:teaser-figure} 
\end{figure}

More pragmatic solutions include only a single sensor and obtain multiple exposures by either spatial (i.e.\,per-pixel varying exposure)~\cite{Heide14, Hajisharif15} or temporal multiplexing (i.e. capturing differently exposed frames)~\cite{Debevec97}.
This simpler hardware setup (and related 
algorithms) has recently seen widespread adoption, and is now found in cameras ranging from professional DSLR to low-cost smartphones.

Early multi-frame exposure fusion algorithms work remarkably well for almost-static scenes (e.g. tripod, reduced motion) but result in ghosting and other motion-related artefacts for dynamic scenes. Various approaches have achieved success in reducing artefacts such as patch-based methods \cite{sen12, zheng13}, noise-based reconstruction \cite{granados13}, sparse correspondences \cite{DBLP:conf/cvpr/gallo15} and image synthesis \cite{hu13} but in recent years, Convolutional Neural Networks (CNNs) have greatly advanced the state-of-the-art for HDR reconstruction, especially for complex dynamic scenes~\cite{kalantari17}.

Most HDR CNNs rely on a rigid setup with a fixed, ordered set of LDR input images, which assumes the medium exposure to be the reference image. The most common mechanism for the merging step is image or feature concatenation, and thus for methods where the feature encoder is not shared among input frames \cite{wu18}, there is a dependency between reference frame choice, relative exposure and input image ordering. Optimal exposure parameters~\cite{Hasinoff10} or fast object motion might constrain the amount of relevant frames available, and in general, broader flexibility in terms of number of frames and choice of reference is necessary to extend applicability without the burden of model tweaking or retraining.

As for frame registration, previous models largely rely on pre-trained or classical \textit{off-the-shelf} optical flow methods that are rarely designed or optimized for the characteristics of exposure-bracketed LDR images. Recent pixel rejection or attention strategies are disconnected from the alignment stage and mostly ignore uncertainty in exposure or motion. 

In this paper, we propose a novel algorithm that addresses these limitations in a holistic and unified way. First, we design a HDR-specific optical flow network which can predict accurate optical flow estimates even when the input and target frames are under- or over-exposed. We do this by using symmetric pooling operations to share information between all $n$ input frames, so any missing information in one frame can be borrowed from other frames. Further, we propose models of exposure and alignment uncertainties which are used by our flow and attention networks to regulate contributions from unreliable and misaligned pixels. Finally we propose a flexible architecture that can process any number of input frames provided in any order.


The contributions of this paper are threefold:
\begin{enumerate}

\item A lightweight HDR-specific optical flow network which can estimate accurate pixel correspondences between LDR frames, even when improperly exposed, by sharing information between all input frames with symmetric pooling operations, and is trained using an HDR-aware self-supervised loss that incorporates exposure uncertainty.
\item Models of exposure and alignment uncertainty which we use to regulate contributions from unreliable and misaligned pixels and greatly reduce ghosting artefacts.
\item  A flexible architecture with a multi-stage fusion mechanism which can estimate an HDR image from an arbitrary set of LDR input images.
\end{enumerate}

\begin{figure*}[t]
\centering
\includegraphics[width=1\textwidth]{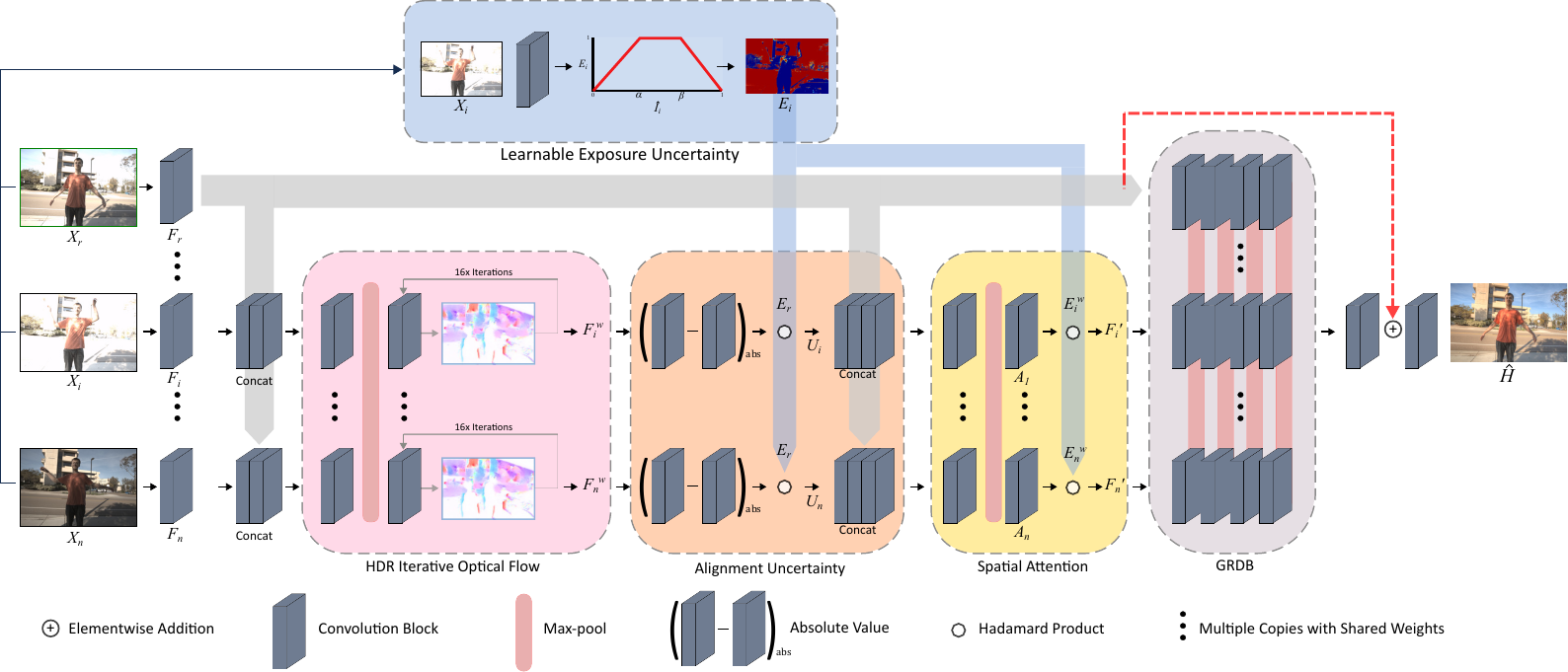}
\caption{Model Architecture. Our model accepts any number of LDR images as input and aligns them with a \emph{HDR flow network} which shares information between frames with pooling operations. We then model exposure and alignment uncertainties which are used by our \emph{attention network} to suppress untrustworthy regions. Finally, the \emph{merging network} consists of a grouped residual dense block with multi-stage max-pooling operations for gradual merging of input frames.}
\label{fig:model-overview}
\end{figure*}


\section{Related Work}
In this section we review the HDR literature with a focus on relevant deep-learning multi-frame exposure fusion methods. For a broader overview we refer the reader to~\cite{Artusi17, Reinhard10}.

The seminal work of \cite{kalantari17} was the first to introduce a training and testing dataset with dynamic scene content. Their proposed method for learning-based HDR fusion is composed of two stages: first, input LDR images are aligned using a classical optical flow algorithm~\cite{liu09} and then a CNN is trained to both merge images and potentially correct any errors in the alignment. Shortly after, \cite{wu18} proposed a similar approach that does not perform a dense optical flow estimation, but rather uses an image-wide homography to perform \textit{background} alignment, leaving the more complex non-rigid \textit{foreground} motions to be handled by the CNN. However, this method is highly dependent on the structure of the reference image, and the magnitude and complexity of the motion. Thus, if certain regions are saturated in the reference image, it fails to accurately reconstruct them in the ﬁnal result. Both \cite{kalantari17} and~\cite{wu18} rely on the optimisation of the HDR reconstruction loss to implicitly learn how to correct ghosting and handle the information coming from different frames. However, neither provides an explicit mechanism to prevent incorrect information (e.g.~overexposed regions) from influencing the final HDR estimation. Despite the noteworthy performance improvement over existing methods at the time, these approaches still suffer from ghosting, especially for fast moving objects and saturated or near-saturated regions.

Yan et al.~\cite{yan19} address some limitations of its predecessors by establishing an attention mechanism to suppress undesired information before the merging stage, e.g.\,misalignments, overexposed regions, and focus instead on desirable details of non-reference frames that might be missing in the reference frame. In the work of Prabhakar et al.~\cite{prabhakar20} parts of the computation, including the optical flow estimation, are performed in a lower resolution and later upscaled back to full resolution using a guide image generated with a simple weight map, thus saving some computation. ~\cite{kalantari19} propose the first end-to-end deeep learning based video HDR algorithm which drastically improved inference speeds compared to classical methods.

More recently, the state of the art in HDR imaging has been pushed to new highs. \cite{niu21} propose the first GAN-based approach to HDR reconstruction which is able to synthesize missing details in areas with disocclusions. Liu et al.~\cite{liu21} introduce a method which uses deformable convolutions as an alignment mechanism instead of optical flow and was the winning submission to the 2021 NTIRE HDR Challenge~\cite{perez21}. Contemporary work has explored and pioneered new training paradigms, such as the weakly supervised training strategy proposed by \cite{prabhakar21}.

Extending these methods to an arbitrary number of images requires changes to the model definition and re-training. Set-processing neural networks \cite{zaheer17} can naturally handle those requirements. In \cite{aittala18}, a permutation invariant CNN is used to deblur a burst of frames which present only rigid, 0-mean translations with no explicit motion registration. 
For the HDR task, \cite{prabhakar19} proposed a method that uses symmetric pooling aggregation to fuse any number of images, but requires pre-alignment~\cite{sun18} and artefact correction by networks which only work on image pairs.


\section{Proposed Method}

Given a set of $n$ LDR images with different exposure values $\left\{ I_{1}, I_{2},\ldots, I_{n} \right\}$ our aim is to reconstruct a single HDR image $H$ which is aligned to a reference frame $I_{r}$. To simplify notation, we denote $I_{r} =I_{1}$, but any input frame can be chosen as the reference frame. To generate the inputs to our model, we follow the work of \cite{yan19, wu18, kalantari17} and form a linearized image $L_{i}$ for each $I_{i}$ as follows:
\begin{equation}
 L_{i} = I_{i}^{\gamma}/t_{i}, 
\end{equation}
where $t_{i}$ is the exposure time of image $I_{i}$ with power-law non-linearity $\gamma$. Setting $\gamma = 2.2$ inverts the CRF, while dividing by the exposure time adjusts all the images to have consistent brightness. We concatenate $I_{i}$ and $L_{i}$ in the channel dimension to form a 6 channel input image $X_{i} = [I_{i},L_{i}]$. Given a set of $n$ inputs $ \left\{ X_{1}, X_{2}, \ldots, X_{n} \right\}$ our proposed network estimates the HDR image $\hat{H}$ by:
\begin{equation}
\hat{H} = h(\{X_{i}\}: \theta), 
\end{equation}
where  $h(\cdot)$ denotes our network, $\theta$ the learned weights of the network and $\hat{H}$ is the predicted radiance map in the linear domain. Our network accepts any number of frames $n$ and is invariant to the order of the non-reference inputs. This is different from the work of \cite{yan19, wu18, kalantari17} where the value of $n$ is fixed to 3 and the order of inputs is fixed, and the work of \cite{prabhakar19} where only the fusion stage is performed on $n$ inputs, but frame alignment and attention are performed on image pairs only. Our method performs alignment, regulates the contribution of each frame based on related alignment and exposure uncertainties and flexibly fuses any number of input frames in a permutation-invariant manner. Our network is also trained end-to-end and $\theta$ is learned entirely during the HDR training.


\subsection{Architecture Overview}
Our architecture is composed of: \textit{Learnable Exposure Uncertainty} (Sec.~\ref{sec:exposure-uncertainty}), \textit{HDR Iterative Optical Flow} (Sec.~\ref{sec:hdr-of}), \textit{Alignment Uncertainty and Attention} (Sec.~\ref{sec:alignment-uncertainty-attention}), and \textit{Merging Network} (Sec.\ref{sec:merging-net}). An overview of the architecture can be seen in Figure~\ref{fig:model-overview}. Our architecture makes use of max-pooling operations to share information between frames and to fuse frames together (Sec.~\ref{sec:set-processing}). This improves the accuracy of our flow and attention networks and gives us the advantage of an architecture that is flexible enough to accept an arbitrary number of images. The flow network and the attention network work together to align non-reference frames to the reference frame and suppress artefacts from misaligned and over-exposed regions. The merging network then combines the aligned features to predict a single HDR image. By explicitly modelling the two most common sources of error, motion and exposure, we create a network that is aware of uncertainty and is able to greatly reduce artefacts compared to state-of-the-art methods, as shown in Figure \ref{fig:teaser-figure}.  

\subsection{Flexible Set Processing}
\label{sec:set-processing}
Many state-of-the-art CNN HDR reconstruction methods require a fixed number of inputs in fixed order of exposure \cite{yan19, wu18, kalantari17}.
To overcome this limitation, we design a set-processing network that can naturally deal with any number of input images. Related concepts have previously shown strong benefits for problems such as deblurring~\cite{aittala18} and we here propose to leverage set-processing and permutation invariance tools for HDR fusion.

Given $n$ input images, our network uses $n$ identical copies of itself with shared weights to process each image separately in its own \textit{stream}. We use a multi-stage fusion mechanism, where features $F_i^k$ of each individual stream $i$ at an arbitrary point $k$ within the network can share information with each other as follows: 
\begin{equation}
\label{eq:local-maxpool}
F_{i}^{\max}=\textrm{conv}(\left[F_{i}^k,\max(F_{1}^k,\ldots,F_{n}^k)\right]),
\end{equation}
where $\max(\cdot)$ denotes a max-pooling operation, $\left[ \cdot \right]$ denotes concatenation and $\textrm{conv}(\cdot)$ denotes a convolutional layer (see Fig.~\ref{fig:maxpool}). This operation is repeated at multiple points in the network. Finally, the  outputs  of  each stream  are  then  pooled  together  into  a  single stream  with a global max-pooling operation $F_\textrm{global}^{\max}=\max(F_{1}^k,\ldots,F_{n}^k)$. This result is processed further in the final layers of our network to obtain the HDR prediction.
This allows the network to process any number of frames in a permutation invariant manner while still being informed by the other frames. 

\begin{figure}[t]
\centering
\includegraphics[width=0.8\columnwidth]{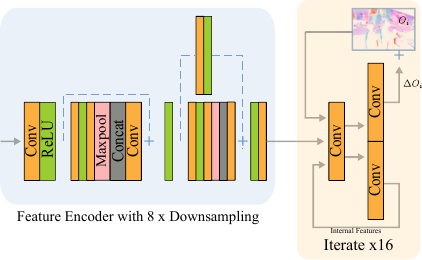}
\caption{Architecture of our HDR Iterative Optical Flow Network. The feature encoder first downsamples the input features by 8x before the recurrent convolutions iteratively refine the estimated optical flow field.}
\label{fig:flow}
\end{figure}

\begin{figure}[t]
\centering
\includegraphics[width=0.6\columnwidth]{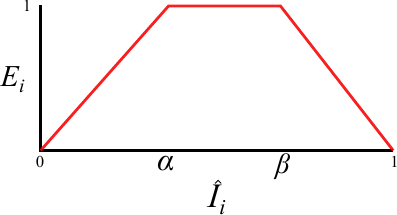}
\caption{We model exposure uncertainty as a piecewise linear function where $\alpha$ and $\beta$ are predicted by the network. Given the mean pixel values of an image, $\hat{I}_i$, our model predicts an image specific response of exposure confidence, $E_i$. }
\label{fig:exposure}
\end{figure}


\subsection{Modelling Exposure Uncertainty}
\label{sec:exposure-uncertainty}

A key limitation of LDR images is that any pixel values above the sensor saturation point results in information loss. Values approaching the saturation level are also unreliable due to negative post-saturation noise~\cite{Hasinoff10}.  When reconstructing an HDR image from multiple LDR images, values at or close to the saturation point can produce artefacts in the final output. Furthermore, underexposed values close to zero are also unreliable due to dark current and low signal-to-noise ratio. We seek to regulate the contribution of such values by modelling our confidence in a given pixel value being correct. For a given input image $I_{i}$, we propose the following piecewise linear function where $\alpha$ and $\beta$ are predicted by the network for each image: 

\begin{equation}
E_{i}  = \begin{cases}
\frac{1}{\alpha}\hat{I_{i}} & \hat{I_{i}} < \alpha \\
1 & \alpha \leqslant \hat{I_{i}} \leqslant \beta \\
\frac{1}{(1 -\beta)}(1 - \hat{I_{i}}) & \beta < \hat{I_{i}}
\end{cases}.
\end{equation}
Here $\hat{I_{i}}$ denotes the mean value across the three RGB channels and $E_{i}$ is the predicted exposure map which represents our estimated confidence in a given pixel. This function is plotted in Figure~\ref{fig:exposure}. We learn from data how to predict $\alpha$ and $\beta$ by means of a shallow network, i.e.~a convolution acting on the concatenation of $[X_{i}, \hat{I_{i}}]$ followed by a spatial average pooling. We constrain $\alpha$ and $\beta$ such that $ 0< \alpha < 0.5 < \beta <1$.  As $\hat{I_{i}}$ approaches 0 or 1, the pixel becomes increasingly unreliable and the value of the exposure mask approaches zero. The slope with which $E_{i}$ approaches zero is determined by $\alpha$ and $\beta$. As shown in Figure~\ref{fig:teaser-figure} this allows us to regulate the contribution that improperly exposed regions in an image can have on our result.

\subsection{HDR Specific Efficient Iterative Optical Flow Network}
\label{sec:hdr-of}

Recent learning based optical flow methods \cite{teed20} typically do not work well for HDR. Successive frames can have large amounts of missing information due to overexposure, which makes aligning frames difficult. This is especially true if the reference and non-reference frames are both overexposed. We solve this issue by using max-pooling operations to share information between all $n$ input frames in our flow network's encoder, as described in Eq.~(\ref{eq:local-maxpool}). This lets the network \textit{fill in} missing information from any of the $n$ available input frames and predict more accurate flows.

The architecture of our proposed flow network is inspired by RAFT \cite{teed20}, however we design the network to be lightweight and efficient. We do not use a context encoder, a correlation layer or a convolutional gated recurrent unit, instead using only simple convolutional layers to predict our optical flow field.

Given an input $X_{i}$ and an exposure mask $E_{i}$, we use a convolutional layer to extract features $F_{i}$ from $X_{i}$. The inputs into the flow network are then concatenated as follows: $\left[ F_{i},F_{r},E_{i} \right]$, where $F_{r}$ corresponds to the features extracted from the reference image. The flow network is informed by $E_{i}$ so that our predictions are aware of the exposure uncertainty in the image. As recurrent convolutions can be computationally expensive at full resolution, the flow network first downsamples the input features by 8$\times$ using strided convolutions. It then iteratively refines the predicted flow over 16 iterations, with a flow initialized to zero, to obtain the optical flow field $O_{i} $ via:

\begin{equation}
     O_{i} = f([F_{i},F_{r},E_{i}]),
\end{equation}
where $f( \cdot )$ denotes our optical flow network. The optical flow field is resized to the original resolution with bilinear upsampling and used to warp our features $F_{i}$:

\begin{equation}
    F_{i}^{w} = w(F_{i},O_{i}),
\end{equation}
where $F_{i}^{w}$ are the warped features and $w( \cdot )$ denotes the function of warping an image with an optical flow field.  The architecture of our flow network can be seen in Figure \ref{fig:flow}.

Unlike other methods which use fixed alignment \cite{wu18, prabhakar19}, our flow network is trained in a fully self-supervised manner. As ground truth optical flows for our datasets are unavailable, we use the self-supervised photometric loss between the reference features $F_{r}$ and the warped features $F_{i}^{w}$ as supervision to guide the learning of the flow network. We multiply the loss by $E_{r}$ so that the reference frame is only used as supervision in regions where it is well exposed. We also apply the optical flow field to the exposure mask, so it remains spatially aligned with the warped features:

\begin{equation}
   E_{i}^{w} = w(E_{i},O_{i}), 
\end{equation}
where $E_{i}^{w}$ is the warped exposure mask.


\begin{figure}[t]
\centering
\includegraphics[width=0.6\columnwidth]{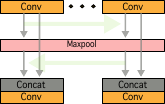}
\caption{An overview of our multi-stage fusion mechanism. The green arrows show the direction of information flow between the different streams. By sharing information between streams at multiple points, the network is able to produce clear, detailed images.}
\label{fig:maxpool}
\end{figure}


\subsection{Alignment Uncertainty and Attention}
\label{sec:alignment-uncertainty-attention}

Our attention network is informed by two measures of uncertainty: exposure and alignment. To model the alignment uncertainty we compute an uncertainty map $U_{i}$ as:
\begin{equation}
U_{i}  =  \mathtt{abs}(F_{i}^{w} -F_{r}) \circ E_{r}, 
\end{equation}
where $\mathtt{abs}(\cdot)$ denotes the elementwise absolute value and $\circ$ denotes element-wise multiplication. This map captures the difference in motion between the reference frame and the warped frame and helps inform our attention network of any inconsistencies in the alignment. We multiply by $E_{r}$ so that only the well exposed regions of the reference frame are used to calculate misalignments. By regulating the contributions from misaligned areas of the image, our network can significantly reduce ghosting in the final output. The exposure uncertainty is given by the warped exposure map $E_{i}^{w}$. The inputs to the attention network are then concatenated as follows [$F_{i}^{w},F_{r},U_{i}, E_{i}^{w}$]. The attention network predicts a 64 channel attention map $ A_{i}$ as follows:

\begin{equation}
 A_{i}= a( [F_{i}^{w},F_{r},U_{i},E_{i}^{w}]), 
\end{equation}
where $a( \cdot )$ denotes our attention network. As in our flow network, we use 
max-pooling to share information between all $n$ input frames. We then obtain our regulated features by multiplying the warped features $F_{i}^{w}$ by the attention map and the exposure map:

\begin{equation}
\label{eqn:final-attention}
F^{'}_{i} =  F_{i}^{w} \circ A_{i} \circ E_{i}^{w}, 
\end{equation}
where $\circ$ denotes element-wise multiplication and $F^{'}_{i}$ denotes the regulated features. Multiplication by the exposure map enforces a strict constraint on our network and 
prevents unreliable information leaking into our output. 
Our HDR-aware attention effectively regulates the contribution of each frame, taking into account both alignment and exposure uncertainty.


\subsection{Merging Network}
\label{sec:merging-net}
Our merging network takes the regulated features obtained from Equation \ref{eqn:final-attention} and merges them into a single HDR image. The merging network is based on a Grouped Residual Dense Block (GRDB) \cite{kim19}, which consists of three Residual Dense Blocks (RDBs) \cite{zhang18}. We modify the GRDB so that each stream can share information with the other streams for a multi-stage fusion of features. An overview of the fusion mechanism can be seen in Figure~\ref{fig:maxpool}. Specifically we add a max-pooling operation after each RDB which follow the formulation described in Equation~\ref{eq:local-maxpool}. This allows the network to progressively merge features from different streams, instead of merging them together in a single concatenation step where information might be lost. This is followed by a final global max-pooling operation which collapses the $n$ streams into one. The merging network then processes this result further with a global residual connection and refinement convolutions.

\subsection{Loss Function}
As HDR images are not viewed in the linear domain, we follow previous work and use the $\mu$-law to map from the linear HDR image to the tonemapped image:
\begin{equation}
\label{eq:mu-law}
\mathcal{T}(H)  = \frac{log(1 + \mu H)}{ log(1 + \mu)}\;,
\end{equation}
where $H$ is the linear HDR image, $\mathcal{T}(H)$ is the tonemapped image and $\mu = 5000$. We then estimate the $\ell_{1}$-norm between the prediction and the ground truth to construct a tone mapped loss as follows:
\begin{equation}
\mathcal{L}_{tm}  = \left\Vert \mathcal{T}(\hat{H}) - \mathcal{T}(H) \right\Vert_{1}\,.
\label{eqn:tonemapped_loss}
\end{equation}

To improve the quality of reconstructed textures we also use the perceptual loss as in~\cite{johnson16}. We pass the tonemapped images through a pre-trained VGG-19~\cite{simonyan15} and extract features from three intermediate layers. We reduce the $\ell_{1}$-norm between the features of the ground truth and our prediction:
\begin{equation}
 \mathcal{L}_{vgg}  = \left\Vert \phi (\mathcal{T}(\hat{H})) - \phi (\mathcal{T}(H)) \right\Vert_{1}\,,
\end{equation}
where $\phi$ is a pre-trained VGG-19 network. Finally, to provide supervision for our optical flow network, we calculate a simple photometric loss between the warped features $F_{i}^{w}$ and the reference features $F_{r}$ and multiply by $E_{r}$ to limit supervision to well exposed regions in the reference frame:

\begin{equation}
\mathcal{L}_{phot} =  \left\Vert (F_{i}^{w} - F_{r}) \circ E_{r}\right\Vert_{1} \;,
\end{equation}
where $\mathtt{abs}$ is the elementwise absolute value.
Our total loss function can be expressed as:
\begin{equation}
 \mathcal{L}_{tot} =  \mathcal{L}_{tm}  + \mathcal{L}_{phot} + 10^{-3}\mathcal{L}_{vgg}\,. 
\end{equation}

\subsection{Implementation details}

During training, we take a random crop of size $256{\times}256$ from the input image. We perform random horizontal and vertical flipping and random rotation by 0\degree,\space 90\degree,\space 180\degree \space or 270\degree \space degrees to further augment the training data. We train using a batch size of $16$ and a learning rate of $0.0001$ with the Adam optimizer. During test time, we run inference on the full test image of size 1500x1000 for the Kalantari et al. dataset, and 1536x813 for the Chen et al. dataset. We implement the model in PyTorch, and train the model on 4 Nvidia V100 GPUs for approximately 2 days.



\begin{figure}[h]
\centering
\includegraphics[width=0.95\columnwidth]{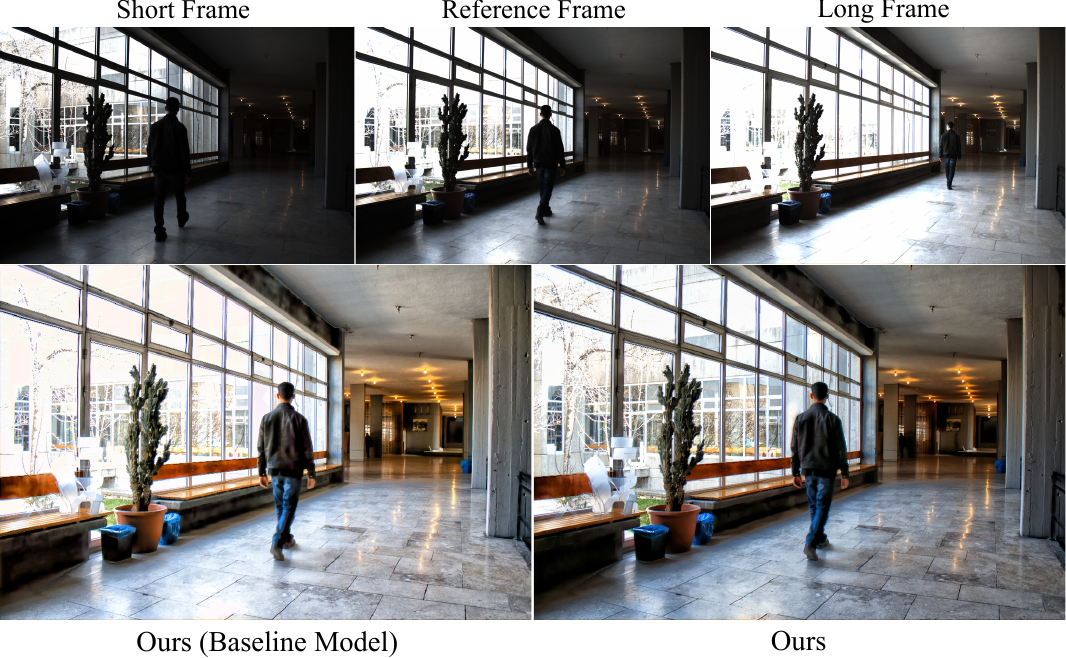}
\caption{Qualitative evaluation of our model on a test scene from the Tursun dataset with large foreground motion and severe over/under-exposure. Our exposure uncertainty is able to regulate the contributions of overexposed pixels in the window panes while our max-pooling mechanism restores details which are not visible in the reference frame.}
\label{fig:tursun-168}
\end{figure}


\begin{figure}[h]
\centering
\includegraphics[width=0.95\columnwidth]{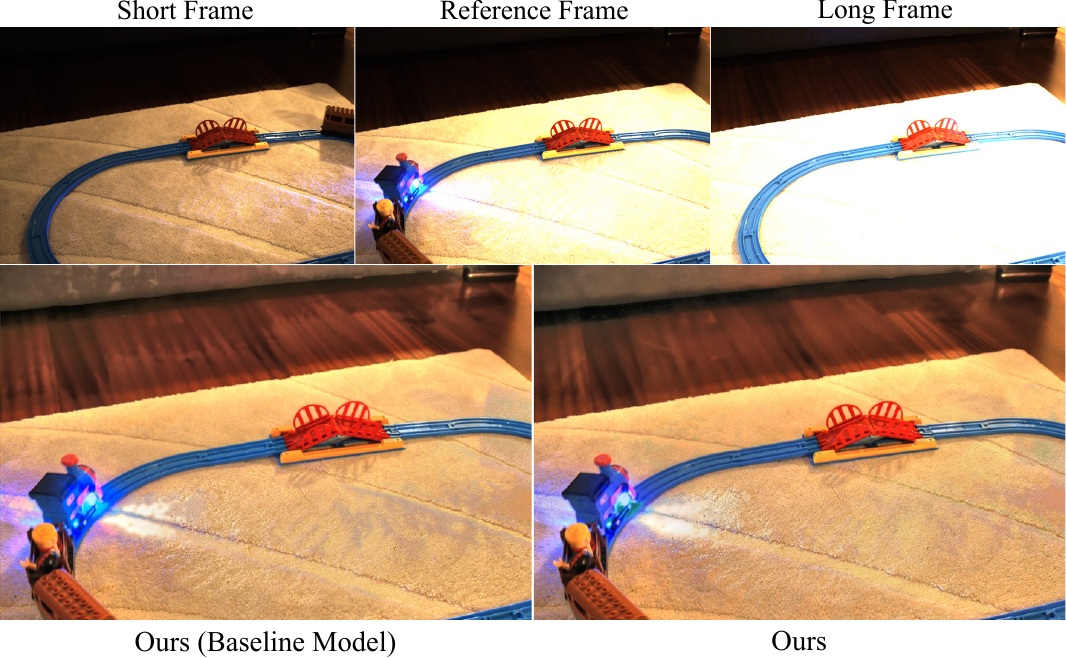}
\caption{Qualitative evaluation of our model on a test scene from the Tursun dataset with a fast moving object. Our efficient flow network reduces ghosting artefacts while our exposure uncertainty reduces exposure artefacts.}
\label{fig:tursun-179}
\end{figure}


\begin{figure}[h]
\centering
\includegraphics[width=0.95\columnwidth]{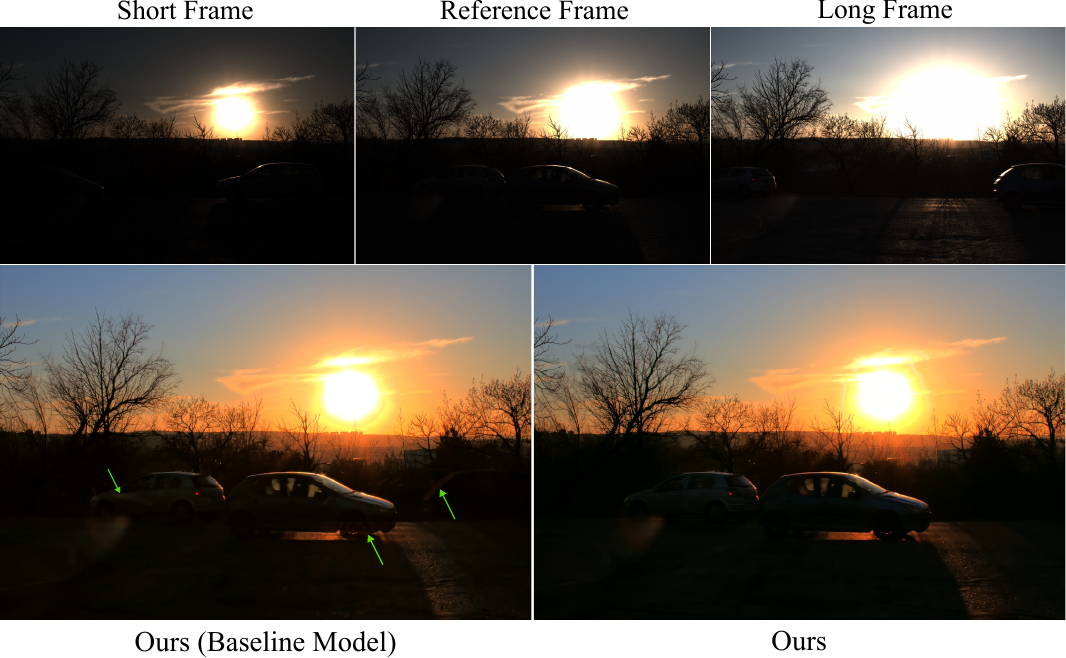}
\caption{Qualitative evaluation of our model on a test scene from the Tursun dataset with multiple fast moving objects which are present in all input frames. The baseline model struggles with ghosting artefacts, indicated by the green arrows. Our method is able to eliminate these artefacts entirely.}
\label{fig:tursun-117}
\end{figure}

\section{Results}


  \begin{table}[t]
\centering
\caption{An ablation study showing the contributions of our proposed fusion mechanism, HDR iterative flow, learnable exposure modelling and alignment uncertainty.}
\begin{adjustbox}{width=0.47\textwidth}
\begin{tabular}{l c c} \toprule
  Model & PSNR-$\mu$ $\pm$ $\textit{\textbf{t}}_{0.95}$  & PSNR-L $\pm$ $\textit{\textbf{t}}_{0.95}$ \\ \midrule
    Baseline Model & 43.91 $\pm$ 0.031 & 41.39 $\pm$ 0.156\\
    + Multi Stage Max-pooling (MSM)  & 44.18 $\pm$ 0.032 & 41.78 $\pm$ 0.172 \\ 
    + MSM + Flow  & 44.23 $\pm$ 0.031 & 42.49 $\pm$  0.178\\ 
    + MSM + Flow + Fixed Exposure  & 44.21 $\pm$ 0.032 & 42.20 $\pm$ 0.168 \\
    + MSM + Flow + Exposure Uncertainty & 44.28 $\pm$ 0.033 & 42.28 $\pm$ 0.194 \\
    + MSM + Flow + Fixed Exposure + Alignment Uncertainty  & 44.28 $\pm$ 0.032 & 42.51 $\pm$ 0.155 \\  
    + MSM + Flow + Exposure Uncertainty + Alignment Uncertainty & \textbf{44.35} $\pm$ 0.033 & \textbf{42.60} $\pm$ 0.165\\ \bottomrule
\end{tabular}
\end{adjustbox}
\label{table:ablation}
 \end{table}

We conduct several experiments both comparing against well-known state-of-the-art algorithms and also individually validating the contributions in an extensive ablation study. The experimental setup is described below.

\textbf{Datasets:} We use the dynamic training and testing datasets provided by Kalantari and Ramamoorthi~\cite{kalantari17} which includes 89 scenes in total. Each of these scenes include three differently exposed input LDR images (with EV of {-2.00, 0.00, +2.00} or {-3.00, 0.00, +3.00}) which contain dynamic elements (e.g.\,camera motion, non-rigid movements) and a ground-truth image aligned with the medium frame captured via static exposure fusion. 
Additionally we use the dynamic testing dataset provided by Chen et al.~\cite{chen21} for further evaluation. As this dataset does not have a corresponding training set, all methods are trained on the Kalantari dataset and evaluated on the Chen dataset. We test on the 3-Exposure setting which has the ground truth aligned to the middle exposure. To keep it consistent with training, we restrict the number of input frames to three with EVs of {-2.00, 0.00, +2.00}.
For purely qualitative evaluation of our method, we include testing sequences from the Tursun \cite{tursun15} dataset.

\textbf{Metrics:} We include seven different objective metrics in our quantitative evaluation. First, we compute the PSNR-L, which is a fidelity metric computed directly on the linear HDR estimations. HDR linear images are normally tonemapped for visualization, and thus we include PSNR-$\mu$, which evaluates PSNR on images tonemapped using the $\mu$-law, as defined in Eq.~(\ref{eq:mu-law}), which is a simple canonical tonemapper. We also calculate PSNR-PU, which uses the perceptual uniform encoding (PU21) for HDR images introduced by \cite{mantiuk21} which aims to improve the correlation between standard metrics and subjective scores on HDR images. For each of the three image domains (linear, $\mu$-tonemapped, PU21), we also calculate the SSIM (Structural Similarity Index) metric introduced by \cite{wang04} which aims to evaluate perceived changes in the underlying structure of the image. This gives us three further metrics, namely SSIM-L, SSIM-$\mu$ and SSIM-PU. Lastly, we also compute the HDR-VDP 2.2~\cite{narwaria15}, which estimates both visibility and quality differences between image pairs. For each metric, we also report a confidence interval calculated using a \textit{t}-test at the 95\% significance level. We compute the confidence intervals per image and report the mean across the test set.


\begin{figure*}[t]
\centering
 \includegraphics[width=0.75\textwidth]{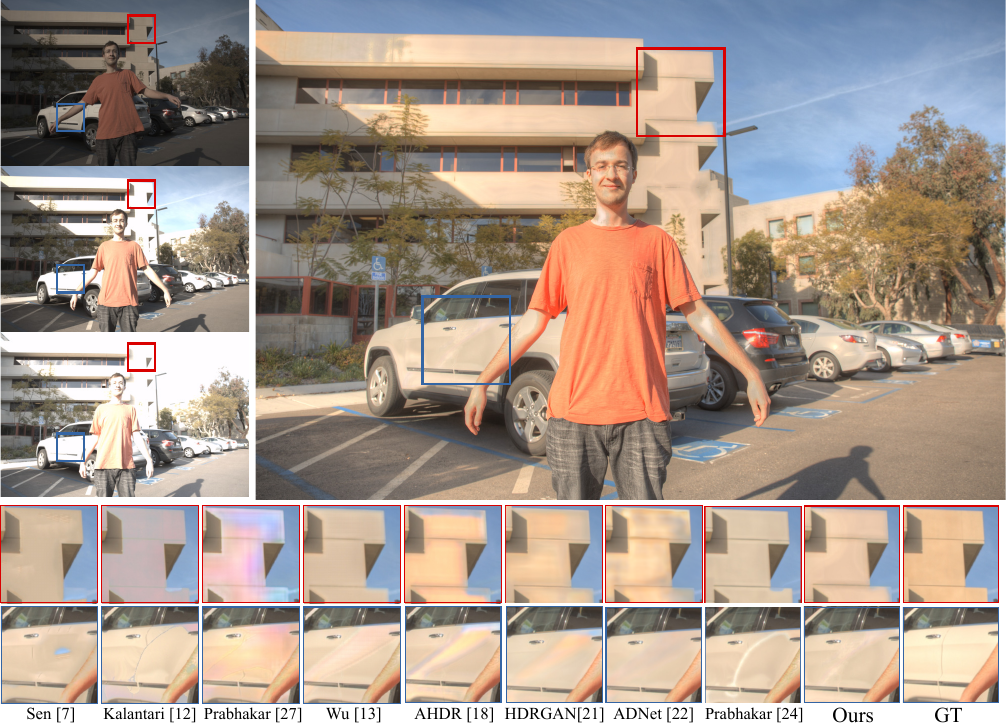}
\caption{Qualitative evaluation of our method for an image from the Kalantari test set. Our method obtains images with noticeably less ghosting artefacts, together with sharper and finer details. Best viewed zoomed-in.}
\label{fig:qualitative-result} 
\end{figure*}


\begin{figure*}[t]
\centering
\includegraphics[width=0.75\textwidth]{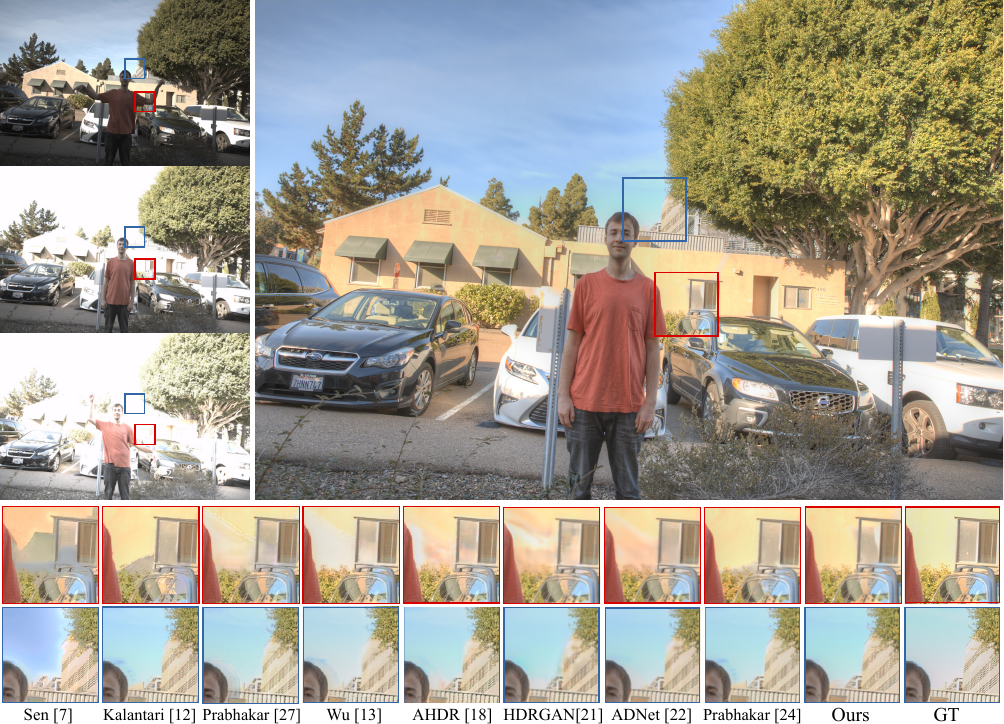} 
\caption{Qualitative evaluation of our method for an image from the Kalantari test set. Our method obtains images with noticeably less ghosting artefacts, together with sharper and finer details. Best viewed zoomed-in.}
\label{fig:qualitative-result-2} 
\end{figure*}

\begin{figure*}[t]
\centering
\includegraphics[width=0.75\textwidth]{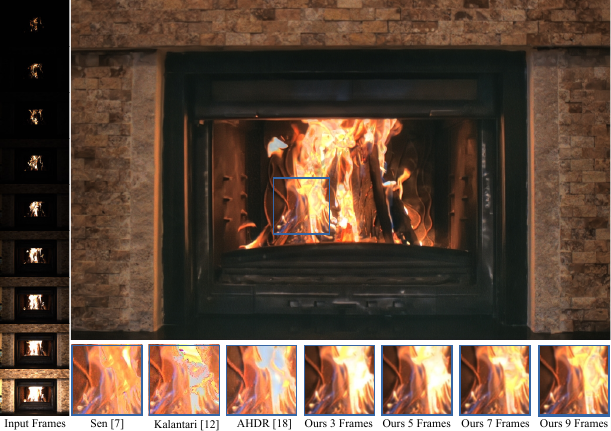}
\caption{ Qualitative evaluation of our model on different numbers of input frames of varying exposures. The quality of the reconstruction clearly increases with the number of input frames used. Our model can utilize information from all 9 input frames, despite having only seen 3 input frames during training.}
\label{fig:tursun-fire}
\end{figure*}


\begin{figure}[h]
\centering
\includegraphics[width=1.0\columnwidth]{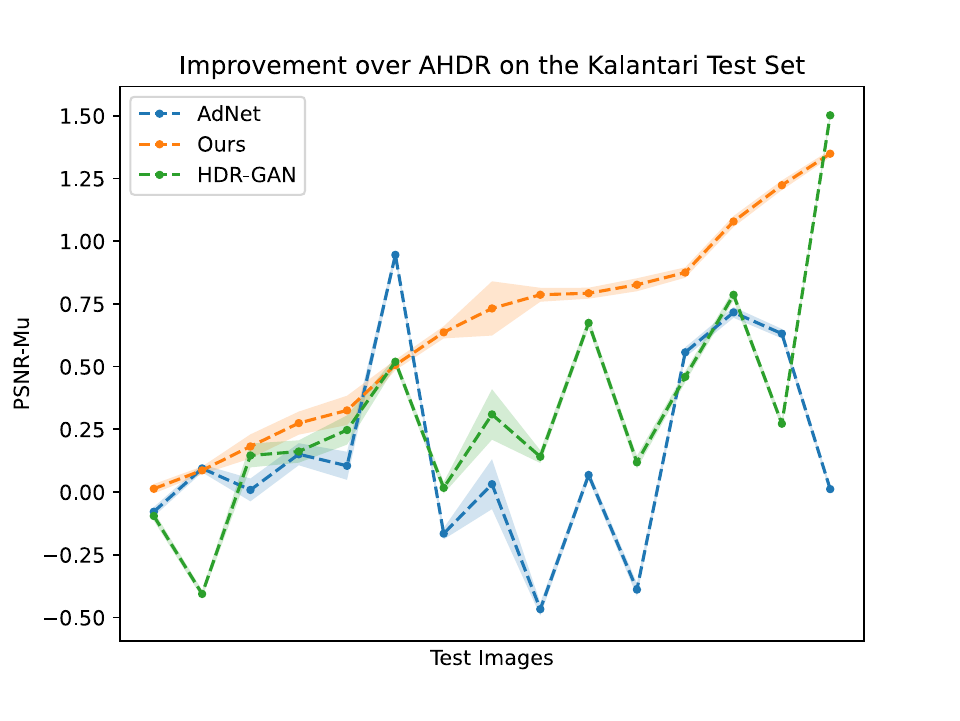}
\caption{We show performance per image on the Kalantari test set of the best performing methods and associated confidence intervals as shaded areas. }
\label{fig:psnr-mu-per-img}
\end{figure}


\begin{figure}[t]
\centering
\includegraphics[width=1.0\columnwidth]{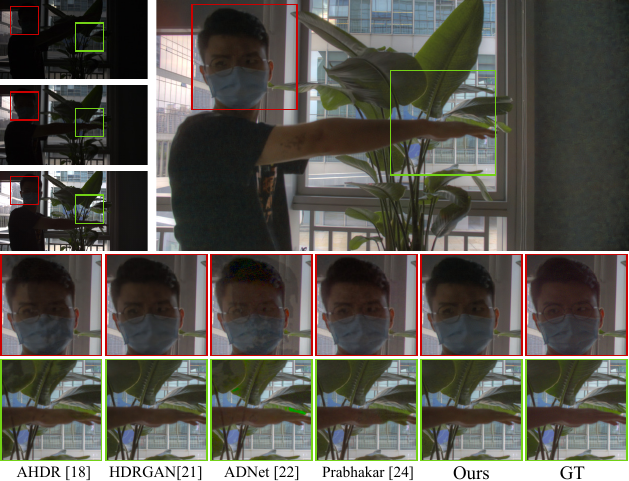} 
\caption{Qualitative evaluation of our method for an image from the Chen test set. Our method performs well even in low light conditions while other deep learning methods suffer from ghosting and artefacts. Best viewed zoomed in.}
\label{fig:qualitative-result-3} 
\end{figure}


\subsection{Ablation Studies}
We evaluate the contribution of the different parts of our model architecture on the Kalantari dataset.

In Table \ref{table:ablation}, we evaluate the quantitative impact of using our multi-stage fusion mechanism as well as the performance gain from our proposed flow network and our uncertainty modelling. Our baseline model uses the same architecture as our proposed method, but with the flow network, uncertainty modelling and multi-stage max-pooling removed, instead using concatenation as the fusion mechanism, and the attention mechanism from \cite{yan19}. We also qualitatively evaluate the impact of our contributions in Figures \ref{fig:tursun-168}, \ref{fig:tursun-179} and \ref{fig:tursun-117}.

\textbf{Fusion Mechanism}. We show in Table~\ref{table:ablation} that using our multi-stage fusion mechanism (MSM) outperforms concatenation (Baseline Model) by 0.39dB PSNR-L and 0.27dB PSNR-$\mu$. The progressive sharing of information between streams allows the network to retain more information and produce sharper, more detailed images.

\textbf{Motion Alignment and Modelling Uncertainty}. We look at the performance of our proposed flow network and uncertainty modelling in Table~\ref{table:ablation}. Our flow network (MSM + Flow)improves PSNR-L by a large 0.7dB, and PSNR-$\mu$ by 0.05dB. compared to using just MSM. We validate the contribution of our learnable model of exposure uncertainty by comparing it to the non-learnable fixed exposure model used by \cite{kalantari17}. We fix the values of $\alpha$ and $\beta$ to match the triangle functions used to generate the ground truths of the Kalantari dataset, which are in essence the oracle $\alpha$ and $\beta$ parameters. Our learnable exposure modelling (MSM + Flow + Exposure Uncertainty) shows an improvement in PSNR-$\mu$ of 0.07dB and PSNR-L of 0.08dB compared to the fixed exposure model (MSM + Flow + Fixed Exposure) In this case, the gain from learning exposure values is small as it is possible to easily fix the values to their optima due to prior knowledge of how the dataset was created. However, this is not always possible, especially in scenarios with different numbers of input images and exposure levels, where the underlying ideal weights are not known. Our decision to learn the weights allows our model to handle any number of input frames without any manual tuning. We also validate the contribution of our alignment uncertainty (MSM + Flow + Fixed Exposure + Alignment Uncertinaty), which gives an improvement in PSNR-$\mu$ of 0.07dB and PSNR-L of 0.32dB when compared to using only exposure uncertainty.


\subsection{Performance Evaluation}
\begin{table*}
\centering
\caption{\label{table:kalantari-results} 
 Quantitative results on the Kalantari et al.~\cite{kalantari17} dataset. Best performer denoted in bold and runner-up in underscored text. $\dagger$ Values as reported by authors.}
\begin{adjustbox}{width=0.99\textwidth}
\begin{tabular}{ l  c  c  c  c  c  c  c  c} \toprule
    Method & PSNR-$\mu$ $\pm$ $\textit{\textbf{t}}_{0.95}$  & PSNR-PU $\pm$ $\textit{\textbf{t}}_{0.95}$ & PSNR-L $\pm$ $\textit{\textbf{t}}_{0.95}$ & SSIM-$\mu$ $\pm$ $\textit{\textbf{t}}_{0.95}$ & SSIM-PU $\pm$ $\textit{\textbf{t}}_{0.95}$ & SSIM-L $\pm$ $\textit{\textbf{t}}_{0.95}$ & HDR-VDP-2 $\pm$ $\textit{\textbf{t}}_{0.95}$\\ \midrule
    Sen \cite{sen12} & 40.98 $\pm$ 0.031 & 33.27 $\pm$ 0.038 & 38.38 $\pm$ 0.140 &  0.9880  $\pm$ \num{4.58e-5} & 0.9782 $\pm$ \num{7.24e-5} & 0.9758 $\pm$ \num{8.37e-5} & 60.54  $\pm$ 1.17 \\
    Kalantari \cite{kalantari17} & 42.70 $\pm$ 0.030 & 33.86 $\pm$ 0.037 & 41.23 $\pm$ 0.146 & 0.9915 $\pm$ \num{3.48e-5} & 0.9832 $\pm$ \num{5.66e-5}&  0.9858 $\pm$ \num{5.85e-5} & 64.63 $\pm$  1.23 \\
    Wu \cite{wu18}& 42.01 $\pm$ 0.024 & 30.82 $\pm$ 0.015 & 41.62 $\pm$ 0.145 & 0.9898 $\pm$ \num{3.13e-5}& 0.9805 $\pm$ \num{5.17e-5} & 0.9872  $\pm$ \num{5.20e-5} & \underline{65.78}  $\pm$ 1.15 \\
    AHDR \cite{yan19} & 43.57 $\pm$  0.031 & 33.46 $\pm$ 0.023 & 41.16 $\pm$ 0.181 & 0.9922 $\pm$ \num{2.98e-5} & 0.9843 $\pm$ \num{5.23e-5} & 0.9871 $\pm$ \num{5.82e-5} & 64.83 $\pm$ 1.19\\
    Prabhakar19 \cite{prabhakar19} & 42.79 $\pm$ 0.027 & 29.06 $\pm$ 0.017 & 40.31 $\pm$ 0.211 & 0.9912 $\pm$ \num{3.08e-5} & 0.9762 $\pm$ \num{5.51e-5}& 0.9874  $\pm$\num{5.44e-5} & 62.95 $\pm$ 1.24 \\
    Pu \cite{pu20}$^\dagger$  & 43.85 & - &  41.65 & 0.9906 & - & 0.9870 &  - \\
    Prabhakar20 \cite{prabhakar20}$^\dagger$ & 43.08 & - & 41.68 & - & - & - & -  \\
    NHDRR \cite{yan20}$^\dagger$  & 42.41 & - & - &  0.9887 & - & - & - \\
    Prabhakar21 \cite{prabhakar21} & 41.94 $\pm$ 0.027 & 32.18 $\pm$ 0.022 & \underline{41.80} $\pm$ 0.141 & 0.9901 $\pm$ \num{3.20e-5} & 0.9813 $\pm$ \num{5.26e-5} & \underline{0.9892}  $\pm$ \num{4.87e-5} & 65.30  $\pm$ 1.15 \\
    ADNet \cite{liu21} & 43.87 $\pm$ 0.031 & 30.68 $\pm$ 0.014 & 41.69 $\pm$ 0.156 & 0.9925 $\pm$ \num{2.88e-5} & 0.9845 $\pm$ \num{4.96e-5} & 0.9885 $\pm$ \num{5.34e-5} &  65.56 $\pm$ 1.14 \\ 
    HDR-GAN \cite{niu21} & \underline{43.96}  $\pm$ 0.032 & \underline{34.04} $\pm$ 0.032& 41.76 $\pm$ 0.164 & \underline{0.9926} $\pm$ \num{2.90e-5} & \underline{0.9853} $\pm$ \num{5.00e-5} & 0.9884 $\pm$ \num{5.43e-5} &   65.07 $\pm$ 1.14\\ \midrule
    Ours  & \textbf{44.35} $\pm$ 0.033 & \textbf{35.13} $\pm$ 0.030 & \textbf{42.60} $\pm$ 0.165 & \textbf{0.9931} $\pm$ \num{2.72e-5}  & \textbf{0.9865} $\pm$ \num{4.57e-5} & \textbf{0.9902} $\pm$ \num{4.61e-5}  &  \textbf{66.56}  $\pm$  1.18\\ \bottomrule
\end{tabular}
\end{adjustbox}

\end{table*}

We evaluate the performance of our proposed method for the HDR estimation task and compare it to other state-of-the-art methods both quantitatively and qualitatively. The methods included in our benchmark cover a broad range of approaches, namely: the patch-based method of Sen et al.~\cite{sen12}, methods which use traditional alignment followed by CNNs to correct dense and global alignment, \cite{kalantari17, prabhakar21, wu18}, the flexible aggregation approach of \cite{prabhakar19} that also uses dense alignment, methods which rely on attention or feature selection followed by a CNN to deghost and merge images \cite{yan19, yan20}, a GAN-based approach which can synthesize missing details in areas with disocclusions~\cite{liu21} and a method which uses deformable convolutions as an alignment mechanism \cite{liu21}. For the Chen et al. test set, we also compare against the HDR video method proposed by \cite{chen21}, which uses a coarse to fine architecture to align and reconstruct input frames. As this method requires seven input frames for the three exposure setting, we do not re-train this on the Kalantari dataset and instead use the pre-trained weights provided by the authors.
We show in Table~\ref{table:kalantari-results} the quantitative evaluation on the Kalantari test set.  The differences in PSNR between our method and the runners up on the Kalantari dataset are large (i.e. +$1.1$dB PSNR-PU, +$0.4$db PSNR-$\mu$, +$0.8$dB PSNR-L). Similarly, the HDR-VDP-2 score obtained by our method outperforms all others by a wide margin (i.e.\,$0.8$). Furthermore, we outperform all methods on all seven metrics, showing our method is consistently better across different evaluation criteria. To further evaluate the consistency of our improvement over other methods, we look at the distribution of PSNRs per image across the Kalantari test set. In Figure \ref{fig:psnr-mu-per-img}, we show the improvement achieved in PSNR-$\mu$ of our method and other top performers over AHDR \cite{yan19}, which is a well known and strong baseline method in HDR imaging. Our method demonstrates a consistent and significant improvement over AHDR, being the only method which achieves an improvement in performance on every single test image. Apart from a few exceptions, our method also outperforms the existing state-of-the-art methods for most images. We observe similar performance on the Chen et al. dynamic test set, demonstrating the generalization ability of our model on out of domain data. Our method is best or second best in six out of seven metrics, with a significant improvement in PSNR-$\mu$ (i.e~$0.5$dB) over the runner up Chen et al. \cite{chen21}. We outperform Chen et al.~on several key metrics such as PSNR-PU, PSNR-$\mu$, SSIM-PU and SSIM-$\mu$ despite the fact their method is trained on video data and has an in-domain advantage.

We also quantitatively evaluate the performance of our optical flow network compared to previous optical flow methods in Table~\ref{table:flow-comparison}. We compare against the traditional method introduced by Liu et al.\cite{liu09} which is used by \cite{kalantari17} and \cite{prabhakar21} in their HDR pipelines, as well as the deep learning based approach introduced by Sun et al.~\cite{sun18} which is used by \cite{prabhakar19} to pre-align input frames. We substitute our optical flow network with the comparison methods but we keep the rest of our proposed architecture the same. We show that our flow network improves on the runner up by 0.55dBs in PSNR-$\mu$ and 0.12dBs in PSNR-L, while having the additional advantages of being end-to-end trainable and requiring no pre-training with ground truth optical flows.

In Figures~\ref{fig:qualitative-result}, \ref{fig:qualitative-result-2}, and \ref{fig:qualitative-result-3} we show some visualizations of our algorithm compared with the benchmarked methods for qualitative, subjective evaluation. All other methods present traces of ghosting artefact around the edges near a moving object, especially where disocclusions happen and one or more frames have overexposed values in those locations (e.g. moving head, moving arm). Our method tackles such challenges effectively thanks to the exposure confidence awareness, and strongly suppresses the ghosting artefact. Additionally, our method also demonstrates better performance when it comes to edges and textures (e.g. building facade), as well as out of domain low-light performance.


\begin{table*}
\centering
\caption{\label{table:video-results} 
Quantitative results on the Chen et al.~\cite{chen21} dynamic dataset. Best performer denoted in bold and runner-up in underscored text. *Chen is trained on a synthetic video dataset while all other methods are trained on Kalantari.}
\begin{adjustbox}{width=0.95\textwidth}
\begin{tabular}{ l  c  c  c  c  c  c  c  c} \toprule
    Method & PSNR-$\mu$ $\pm$ $\textit{\textbf{t}}_{0.95}$  & PSNR-PU $\pm$ $\textit{\textbf{t}}_{0.95}$ & PSNR-L $\pm$ $\textit{\textbf{t}}_{0.95}$ & SSIM-$\mu$ $\pm$ $\textit{\textbf{t}}_{0.95}$ & SSIM-PU $\pm$ $\textit{\textbf{t}}_{0.95}$ & SSIM-L $\pm$ $\textit{\textbf{t}}_{0.95}$ & HDR-VDP-2 $\pm$ $\textit{\textbf{t}}_{0.95}$\\ \midrule
    Sen \cite{sen12} & 40.79 $\pm$ 0.011 & \textbf{34.33} $\pm$ 0.012 & 39.58 $\pm$ 0.043 & 0.9862 $\pm$ \num{2.35e-5} & 0.9617 $\pm$ \num{6.41e-5} & 0.9912 $\pm$ \num{5.86e-5} &  68.83 $\pm$ 0.77 \\
    AHDR \cite{yan19} & 39.56 $\pm$ 0.019 & 26.71 $\pm$ 0.006 & 35.09 $\pm$ 0.058 & 0.9814 $\pm$ \num{3.84e-5} & 0.9632 $\pm$ \num{7.08e-5} & 0.9900 $\pm$ \num{7.24e-5}& 63.78 $\pm$ 0.75\\
    Prabhakar21 \cite{prabhakar21} &  40.89 $\pm$ 0.015 & 29.06 $\pm$ 0.008 & 38.19 $\pm$ 0.052 & 0.9857 $\pm$ \num{3.17e-5} & 0.9670 $\pm$ \num{6.16e-5}& 0.9899 $\pm$ \num{6.27e-5} & 65.48 $\pm$ 0.69 \\
    ADNet \cite{liu21} & 38.14 $\pm$ 0.042 &  30.45 $\pm$ 0.029 & 40.97 $\pm$ 0.042 & 0.9797 $\pm$ \num{4.30e-5} & 0.9622 $\pm$ \num{7.14e-5} & \underline{0.9935} $\pm$ \num{4.15e-5} & \underline{70.00}  $\pm$ 0.66\\ 
    HDR-GAN \cite{niu21} & 40.80 $\pm$ 0.018 & 27.37 $\pm$ 0.007 &  36.19 $\pm$ 0.048 & \textbf{0.9904} $\pm$ \num{2.06e-5} & \underline{0.9730} $\pm$ \num{5.28e-5} & 0.9907 $\pm$ \num{6.37e-5} & 66.33 $\pm$ 0.72\\  
    Chen* \cite{chen21} & \underline{41.47} $\pm$ 0.014 & 33.37 $\pm$ 0.011 & \textbf{42.33} $\pm$ 0.038 & 0.9883 $\pm$ \num{2.65e-5} & 0.9698 $\pm$ \num{5.35e-5} & \textbf{0.9945} $\pm$ \num{3.40e-5} & \textbf{71.87} $\pm$ 0.82 \\ \midrule
    Ours  & \textbf{41.98} $\pm$ 0.014 & \underline{33.67} $\pm$ 0.011 &  \underline{41.32} $\pm$ 0.053 & \underline{0.9884} $\pm$ \num{2.41e-5} & \textbf{0.9732} $\pm$  \num{4.84e-5} & \underline{0.9935} $\pm$ \num{4.08e-5}& 69.42 $\pm$ 0.78\\ \bottomrule
\end{tabular}
\end{adjustbox}

\end{table*}



\begin{figure*}[!tp]
\centering
\includegraphics[width=0.85\textwidth]{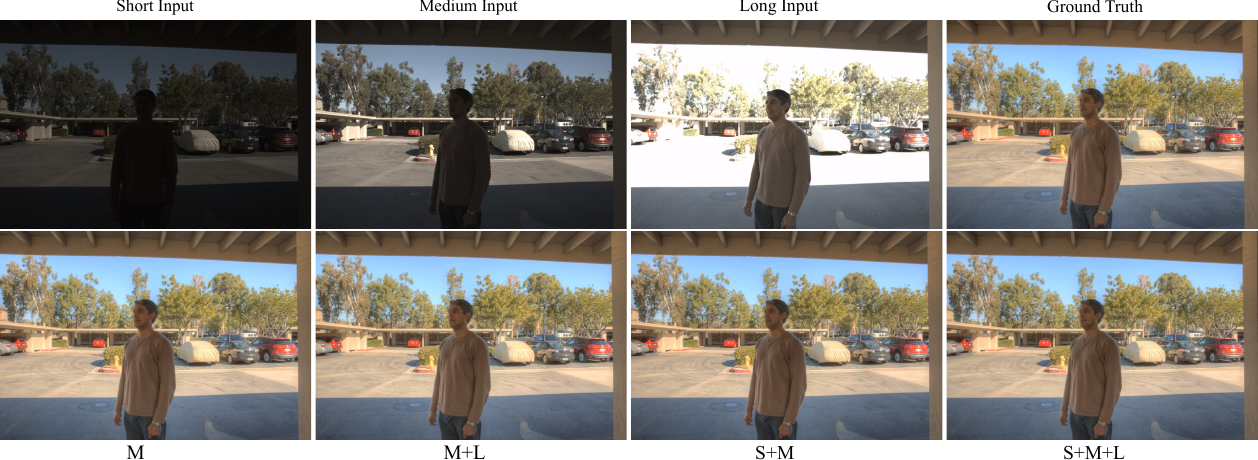}
\caption{Qualitative comparison of our model trained on all permutations with different input frames. When the medium frame is well exposed, our model can attain a high quality prediction with just one frame. There is no noticeable increase in image quality when more inputs are provided. }
\label{fig:permutations-well-exposed} 
\end{figure*}
\begin{figure*}[!bp]
\centering
\includegraphics[width=0.85\textwidth]{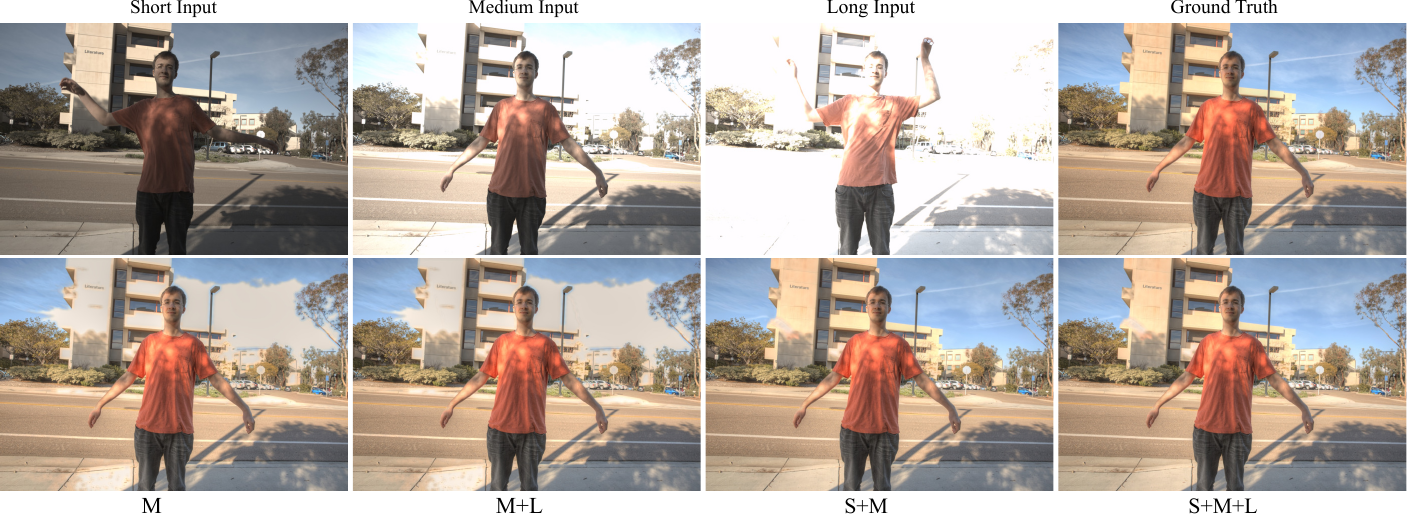}
\caption{Qualitative comparison of our model trained on all permutations with different input frames. When the medium frame is overexposed, our model is not able to completely hallucinate details in large overexposed regions. The short frame is essential to reconstruct the missing details. There is no noticeable improvement in image quality from including the extremely overexposed long frame.}
\label{fig:permutations-over-exposed} 
\end{figure*}

\begin{figure}[!p]
\centering
\includegraphics[width=0.9\columnwidth]{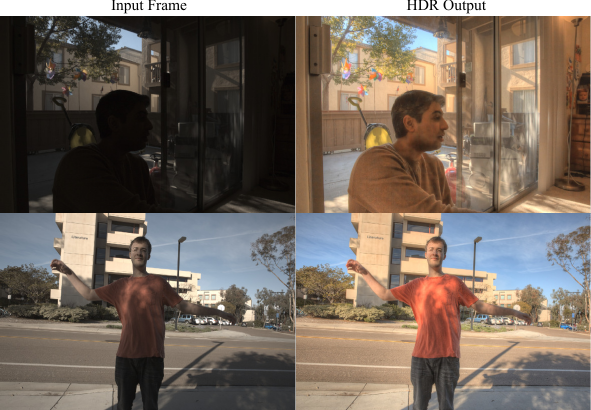}
\caption{Qualitative evaluation of our model using just the short frame as input. When the input frame is severely underexposed (top), we see quantization and noise related artefacts. However when the input frame is well exposed (bottom), the output is free of artefacts.}
\label{fig:short-exposure}
\end{figure}
  \begin{figure}[!p]
\centering
\includegraphics[width=0.9\columnwidth]{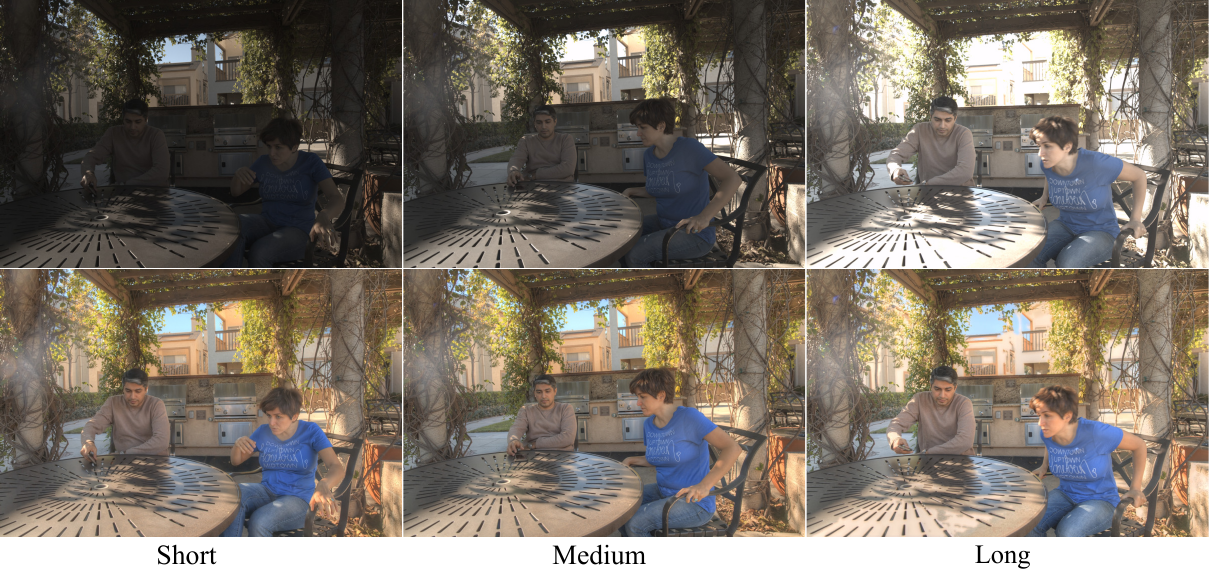}
\caption{
Our method is flexible enough to accept any frame as the reference frame without re-training, providing superior choice to the user.}
\label{fig:flexible-figure-1} 
\end{figure}


\begin{table}
\centering
\caption{Comparison of different training regimes on the Kalantari test set using our proposed model.
* Batches sampled uniformly from the two options.
** Batches sampled uniformly from all four options. The four options are: S + M + L, S + M, M + L, M.}
\begin{adjustbox}{width=0.47\textwidth}
\begin{tabular}{c c c c} \toprule
    Training Frames &  Test Frames  & PSNR-$\mu$  $\pm$ $\textit{\textbf{t}}_{0.95}$  & PSNR-L  $\pm$ $\textit{\textbf{t}}_{0.95}$  \\ \midrule    
    S + M + L &  S + M + L &  \textbf{44.35} $\pm$ 0.033 &  \textbf{42.60} $\pm$ 0.165 \\ 
    S + M / M + L* &  S + M + L & 43.44 $\pm$  0.031 & 40.58 $\pm$  0.144 \\
    M & S + M + L & 20.86 $\pm$ 0.006 & 24.66 $\pm$ 0.030 \\
    All Permutations ** &   S + M + L &   \underline{44.24} $\pm$ 0.031 &  \underline{42.30} $\pm$ 0.170 \\ \midrule
    S + M + L &  S + M  & 40.24 $\pm$  0.028 &  41.15 $\pm$  0.158\\
    S + M / M + L* &  S + M  &  \underline{44.11} $\pm$ 0.032 &  \underline{42.17} $\pm$ 0.167 \\
    M & S + M &  22.68 $\pm$ 0.005  & 25.38 $\pm$ 0.037 \\
    All Permutations ** &   S + M  & \textbf{44.18} $\pm$  0.031 & \textbf{42.29} $\pm$ 0.168\\ \midrule
    S + M + L &  M + L & 41.85 $\pm$  0.025 &  37.60 $\pm$ 0.148 \\
    S + M / M + L* &  M + L & \underline{42.58} $\pm$ 0.031 &  \underline{38.32} $\pm$  0.234\\
    M & M + L & 22.03 $\pm$ 0.006 & 25.00 $\pm$ 0.032\\
    All Permutations ** &   M + L & \textbf{42.74} $\pm$ 0.030 &  \textbf{38.38} $\pm$ 0.222\\ \midrule
    S + M + L &  M &  33.10 $\pm$ 0.010 & 32.84 $\pm$ 0.032 \\
    S + M / M + L* &  M & 28.26 $\pm$ 0.005 & 22.44 $\pm$ 0.019\\
    M & M &  \textbf{42.86} $\pm$ 0.031 & \textbf{38.79} $\pm$ 0.226 \\
    All Permutations ** &  M & \underline{42.67} $\pm$ 0.030 &  \underline{38.40} $\pm$ 0.216\\ \bottomrule
\end{tabular}
\end{adjustbox}

\label{table:permutation}
\end{table}


\begin{table}[h]
\centering
\caption{Quantitative comparison of different optical flow methods on the Kalantari test set.}
\begin{adjustbox}{width=0.3\textwidth}
\begin{tabular}{c c c c} \toprule 
    Method & PSNR-$\mu$  $\pm$ $\textit{\textbf{t}}_{0.95}$  & PSNR-L  $\pm$ $\textit{\textbf{t}}_{0.95}$  \\ \midrule    
    Liu et al.~\cite{liu09} &  43.68 $\pm$ 0.030 & \underline{42.48} $\pm$ 0.157\\
    PWC-Net.~\cite{sun18}  &  \underline{43.80} $\pm$  0.031 & 42.36 $\pm$ 0.143 \\ \midrule
    Ours  &  \textbf{44.35} $\pm$ 0.033 &  \textbf{42.60} $\pm$ 0.165\\ \bottomrule
\end{tabular}
\end{adjustbox}
\label{table:flow-comparison}
\end{table}


\subsection{Flexible Imaging}
 We show that our model is flexible enough to accept an arbitrary number of images without the need for re-training. In Table \ref{table:permutation} we evaluate the performance of our proposed model when trained and tested on different numbers of images with different exposures. We use the following input frame configurations for training and testing: Short + Medium + Long (S + M + L), Short + Medium (S + M), Medium + Long (M + L), Medium (M). The reference frame in all settings is the medium frame, which is spatially aligned to the ground truth. As expected, performance is best when the testing configuration is seen during training. Our model trained on all permutations achieves competitive cross-setting performance, obtaining the best results for the S + M and M + L settings. It is also competitive with our best model for the M and S + M + L settings, without needing any extra training time, and is capable of accepting a range of different input configurations without the need for re-training. In fact, we show that our model using only two frames (S + M) can obtain results outperforming current state-of-the-art methods using all three frames in PSNR-$\mu$ \cite{niu21} and PSNR-L \cite{prabhakar21}. We qualitatively show the performance gains from our flexible architecture in Figure \ref{fig:tursun-fire}. The figure shows that our method works with 3, 5, 7, or 9 input frames with varying exposures from -4.00 to +4.00, without the need to re-train. As the number of frames increases, the reconstruction quality also clearly increases (i.e. the colour and detail of the flames improves as we go from 3 to 9 frames). Our proposed method is capable of utilizing the information provided in all 9 input frames. In Figure \ref{fig:permutations-well-exposed} we show that our model is also capable of producing high quality HDR outputs with fewer than 3 input frames. Furthermore we show in Figure \ref{fig:flexible-figure-1} that our method can use any frame as the reference frame without re-training, providing superior flexibility and choice when compared to state-of-the-art methods.

 \subsection{Parameters and Runtime}

 We provide a breakdown of our model parameters by sub-components in Table~\ref{table:model-param-bd} and provide a comparison of our flow network with state-of-the-art optical flow methods in Table~\ref{table:model-param-flow}. Our flow network is an order of magnitude smaller than \cite{sun18}, which is used by \cite{prabhakar19} to align images, and 6$\times$ smaller than RAFT \cite{teed20}. We also explore how the runtime of our model varies depending on both the input resolution and the number of input frames in Table~\ref{table:runtimes}. The model runtime grows approximately linearly with the number of input frames, and quadratically with the input resolution.
 

\begin{table}
\centering
\caption{\label{table:runtimes} A comparison of runtimes of our method on different input resolutions and number of input frames, computed on an Nvidia V100 GPU.}
\begin{adjustbox}{width=0.3\textwidth}
\begin{tabular}{ c  c  c} \toprule
    Input Resolution & \# Input Frames & Runtime (s) \\ \midrule
    1024x682 & 1 & 0.24\\
    1024x682 & 2 & 0.48 \\
    1024x682 & 3 & 0.70 \\
    1280x720 & 1 & 0.32\\
    1280x720 & 2 & 0.58\\
    1280x720 & 3 & 0.92 \\
    1500x1000 & 1 & 0.51 \\
    1500x1000 & 2 & 0.97 \\
    1500x1000 & 3 & 1.55 \\ \bottomrule
    \end{tabular}
\end{adjustbox}
\end{table}


 \subsection{Limitations}
One limitation of our method is that it is not able to hallucinate details in large over-exposed regions, as seen in Figure \ref{fig:permutations-over-exposed}. The missing information needs to be provided in one of the input frames for our model to be able to accurately reconstruct the HDR image. This is not surprising given that the loss functions used during training have a focus on HDR reconstruction. There is potential to improve our single image HDR performance by exploring a training strategy similar to those used for inpainting. Similarly our model can not fully denoise extremely underexposed regions, as shown in Figure \ref{fig:short-exposure}. We also note that for the datasets used in our work, the CRF is a simple fixed gamma curve with $\gamma = 2.2$, as defined by the dataset authors. However in a general case, the gamma correction we use is only a coarse approximation of the CRF and we rely on the network to implicitly perform finer adjustments. We expect accurate and explicit estimation of the CRF \cite{grossberg04, lee12} to positively impact the reconstruction performance of our model, especially for cases where the CRF is non-trivial.  Finally, although our model can theoretically accept any number of input frames, the amount of activation memory required increases linearly with the number of input frames. On a single Nvidia V100 GPU, we can process up to nine full sized input frames from the Tursun dataset as shown in Figure \ref{fig:tursun-fire}.


\begin{table}
\centering
\caption{\label{table:model-param-bd} Breakdown of our model parameters by model sub-components.}
\begin{adjustbox}{width=0.3\textwidth}
\begin{tabular}{  c  c  c  c} \toprule
    Model &  \# Parameters (millions) \\ \midrule
    Flow Network & 0.87 \\
    Attention Network & 0.33 \\
    Merging Network & 0.92 \\  \midrule
    Ours Total & 2.12 \\ \bottomrule

    \end{tabular}
\end{adjustbox}
\end{table}

 
\begin{table}
\centering
\caption{\label{table:model-param-flow} A comparison of the number of parameters in our efficient flow network and other state-of-the-art flow networks.}
\begin{adjustbox}{width=0.3\textwidth}
\begin{tabular}{  c  c  c  c} \toprule
    Model & \# Parameters (millions) \\ \midrule
    PWC-Net \cite{sun18} & 8.75 \\
    PWC-Net-Small \cite{sun18} & 4.08 \\
    RAFT \cite{teed20} & 5.30 \\\midrule
    Our Flow Network & 0.87 \\ \bottomrule
    \end{tabular}
\end{adjustbox}
\end{table}


\section{Conclusion}
In this paper we explored modelling exposure and alignment uncertainties to improve HDR imaging performance. We presented (1) an HDR-specific optical flow network which is capable of accurate flow estimations, even with improperly exposed input frames, by sharing information between input images with a symmetric pooling operation. (2) We also presented models of exposure and alignment uncertainty which we use to regulate contributions from unreliable and misaligned pixels and greatly reduce ghosting artefacts. (3) Lastly a flexible architecture which uses a multi-stage fusion to estimate an HDR image from an arbitrary set of LDR input images. We conducted extensive ablation studies where we validate individually each of our contributions. We compared our method to other state-of-the-art algorithms obtaining significant improvements for all the measured metrics and noticeably improved visual results.

\bibliography{tip-bib}

\begin{thebibliography}{10}
\providecommand{\url}[1]{#1}
\csname url@samestyle\endcsname
\providecommand{\newblock}{\relax}
\providecommand{\bibinfo}[2]{#2}
\providecommand{\BIBentrySTDinterwordspacing}{\spaceskip=0pt\relax}
\providecommand{\BIBentryALTinterwordstretchfactor}{4}
\providecommand{\BIBentryALTinterwordspacing}{\spaceskip=\fontdimen2\font plus
\BIBentryALTinterwordstretchfactor\fontdimen3\font minus
  \fontdimen4\font\relax}
\providecommand{\BIBforeignlanguage}[2]{{%
\expandafter\ifx\csname l@#1\endcsname\relax
\typeout{** WARNING: IEEEtran.bst: No hyphenation pattern has been}%
\typeout{** loaded for the language `#1'. Using the pattern for}%
\typeout{** the default language instead.}%
\else
\language=\csname l@#1\endcsname
\fi
#2}}
\providecommand{\BIBdecl}{\relax}
\BIBdecl

\bibitem{Debevec97}
P.~E. Debevec and J.~Malik, ``Recovering high dynamic range radiance maps from
  photographs,'' in \emph{ACM SIGGRAPH 2008 Classes}, 2008.

\bibitem{Tocci11}
M.~D. Tocci, C.~Kiser, N.~Tocci, and P.~Sen, ``A versatile hdr video production
  system,'' in \emph{SIGGRAPH}, 2011.

\bibitem{McGuire07}
M.~{McGuire}, W.~{Matusik}, H.~{Pfister}, B.~{Chen}, J.~F. {Hughes}, and S.~K.
  {Nayar}, ``Optical splitting trees for high-precision monocular imaging,''
  \emph{IEEE Computer Graphics and Applications}, vol.~27, no.~2, 2007.

\bibitem{Froehlich14}
J.~Froehlich, S.~Grandinetti, B.~Eberhardt, S.~Walter, A.~Schilling, and
  H.~Brendel, ``Creating cinematic wide gamut hdr-video for the evaluation of
  tone mapping operators and hdr-displays,'' in \emph{Proc. of SPIE Electronic
  Imaging}, 2014.

\bibitem{Heide14}
F.~Heide, M.~Steinberger, Y.-T. Tsai, M.~Rouf, D.~Pajak, D.~Reddy, O.~Gallo,
  J.~Liu, W.~Heidrich, K.~Egiazarian, J.~Kautz, and K.~Pulli, ``Flex{ISP}: A
  flexible camera image processing framework,'' \emph{ACM Trans. Graph.},
  vol.~33, no.~6, 2014.

\bibitem{Hajisharif15}
S.~Hajisharif and J.~Kronander, J.and~Unger, ``Adaptive dualiso hdr
  reconstruction,'' \emph{EURASIP Journal on Image and Video Processing},
  no.~41, 2015.

\bibitem{sen12}
P.~Sen, N.~K. Kalantari, M.~Yaesoubi, S.~Darabi, D.~B. Goldman, and
  E.~Shechtman, ``Robust patch-based hdr reconstruction of dynamic scenes.''
  \emph{ACM Trans. Graph.}, vol.~31, no.~6, pp. 203--1, 2012.

\bibitem{zheng13}
J.~Zheng, Z.~Li, Z.~Zhu, S.~Wu, and S.~Rahardja, ``Hybrid patching for a
  sequence of differently exposed images with moving objects,'' \emph{IEEE
  Transactions on Image Processing}, vol.~22, no.~12, pp. 5190--5201, 2013.

\bibitem{granados13}
\BIBentryALTinterwordspacing
M.~Granados, K.~I. Kim, J.~Tompkin, and C.~Theobalt, ``Automatic noise modeling
  for ghost-free {HDR} reconstruction,'' \emph{{ACM} Trans. Graph.}, vol.~32,
  no.~6, pp. 201:1--201:10, 2013. [Online]. Available:
  \url{https://doi.org/10.1145/2508363.2508410}
\BIBentrySTDinterwordspacing

\bibitem{DBLP:conf/cvpr/gallo15}
\BIBentryALTinterwordspacing
O.~Gallo, A.~J. Troccoli, J.~Hu, K.~Pulli, and J.~Kautz, ``Locally non-rigid
  registration for mobile {HDR} photography,'' in \emph{2015 {IEEE} Conference
  on Computer Vision and Pattern Recognition Workshops, {CVPR} Workshops 2015,
  Boston, MA, USA, June 7-12, 2015}.\hskip 1em plus 0.5em minus 0.4em\relax
  {IEEE} Computer Society, 2015, pp. 48--55. [Online]. Available:
  \url{https://doi.org/10.1109/CVPRW.2015.7301366}
\BIBentrySTDinterwordspacing

\bibitem{hu13}
\BIBentryALTinterwordspacing
J.~Hu, O.~Gallo, K.~Pulli, and X.~Sun, ``{HDR} deghosting: How to deal with
  saturation?'' in \emph{2013 {IEEE} Conference on Computer Vision and Pattern
  Recognition, Portland, OR, USA, June 23-28, 2013}.\hskip 1em plus 0.5em minus
  0.4em\relax {IEEE} Computer Society, 2013, pp. 1163--1170. [Online].
  Available: \url{https://doi.org/10.1109/CVPR.2013.154}
\BIBentrySTDinterwordspacing

\bibitem{kalantari17}
N.~K. Kalantari and R.~Ramamoorthi, ``Deep high dynamic range imaging of
  dynamic scenes,'' \emph{ACM Transactions on Graphics (Proceedings of SIGGRAPH
  2017)}, vol.~36, no.~4, 2017.

\bibitem{wu18}
S.~Wu, J.~Xu, Y.~Tai, and C.~Tang, ``End-to-end deep {HDR} imaging with large
  foreground motions,'' in \emph{European Conference on Computer Vision}, 2018.

\bibitem{Hasinoff10}
S.~W. {Hasinoff}, F.~{Durand}, and W.~T. {Freeman}, ``Noise-optimal capture for
  high dynamic range photography,'' in \emph{IEEE Conference on Computer Vision
  and Pattern Recognition}, 2010.

\bibitem{Artusi17}
T.~E. Alessandro~Artusi, Thomas~Richter and R.~K. Mantiuk, ``High dynamic range
  imaging technology,'' \emph{IEEE Signal Processing Magazine}, vol.~34, no.~5,
  2017.

\bibitem{Reinhard10}
E.~Reinhard, W.~Heidrich, P.~Debevec, S.~Pattanaik, G.~Ward, and K.~Myszkowski,
  \emph{High Dynamic Range Imaging}, 2nd~ed.\hskip 1em plus 0.5em minus
  0.4em\relax Morgan Kaufmann, 2010.

\bibitem{liu09}
C.~Liu, ``Beyond pixels: Exploring new representations and applications for
  motion analysis,'' Ph.D. dissertation, Massachusetts Institute of Technology,
  2009.

\bibitem{yan19}
Q.~Yan, D.~Gong, Q.~Shi, A.~van~den Hengel, C.~Shen, I.~D. Reid, and Y.~Zhang,
  ``Attention-guided network for ghost-free high dynamic range imaging,'' in
  \emph{Computer Vision and Pattern Recognition}, 2019.

\bibitem{prabhakar20}
\BIBentryALTinterwordspacing
K.~R. Prabhakar, S.~Agrawal, D.~K. Singh, B.~Ashwath, and R.~V. Babu, ``Towards
  practical and efficient high-resolution {HDR} deghosting with {CNN},'' in
  \emph{European Conference Computer Vision}, ser. Lecture Notes in Computer
  Science, vol. 12366.\hskip 1em plus 0.5em minus 0.4em\relax Springer, 2020,
  pp. 497--513. [Online]. Available:
  \url{https://doi.org/10.1007/978-3-030-58589-1\_30}
\BIBentrySTDinterwordspacing

\bibitem{kalantari19}
\BIBentryALTinterwordspacing
N.~K. Kalantari and R.~Ramamoorthi, ``Deep {HDR} video from sequences with
  alternating exposures,'' \emph{Comput. Graph. Forum}, vol.~38, no.~2, pp.
  193--205, 2019. [Online]. Available: \url{https://doi.org/10.1111/cgf.13630}
\BIBentrySTDinterwordspacing

\bibitem{niu21}
Y.~Niu, J.~Wu, W.~Liu, W.~Guo, and R.~W.~H. Lau, ``Hdr-gan: Hdr image
  reconstruction from multi-exposed ldr images with large motions,'' \emph{IEEE
  Transactions on Image Processing}, vol.~30, pp. 3885--3896, 2021.

\bibitem{liu21}
Z.~Liu, W.~Lin, X.~Li, Q.~Rao, T.~Jiang, M.~Han, H.~Fan, J.~Sun, and S.~Liu,
  ``Adnet: Attention-guided deformable convolutional network for high dynamic
  range imaging,'' in \emph{2021 IEEE/CVF Conference on Computer Vision and
  Pattern Recognition Workshops (CVPRW)}, 2021, pp. 463--470.

\bibitem{perez21}
E.~Pérez-Pellitero, S.~Catley-Chandar, A.~Leonardis, R.~Timofte, X.~Wang,
  Y.~Li, T.~Wang, F.~Song, Z.~Liu, W.~Lin, X.~Li, Q.~Rao, T.~Jiang, M.~Han,
  H.~Fan, J.~Sun, S.~Liu, X.~Chen, Y.~Liu, Z.~Zhang, Y.~Qiao, C.~Dong, E.~Y.
  Lyn~Chee, S.~Shen, Y.~Duan, G.~Chen, M.~Sun, Y.~Gao, L.~Zhang, A.~K. A, J.~C.
  V, S.~M.~A. Sharif, R.~A. Naqvi, M.~Biswas, S.~Kim, C.~Xia, B.~Zhao, Z.~Ye,
  X.~Lu, Y.~Cao, J.~Yang, Y.~Cao, G.~R.~K. S, S.~Deepak~Lomte, N.~Krishnan, and
  B.~H. Pawan~Prasad, ``Ntire 2021 challenge on high dynamic range imaging:
  Dataset, methods and results,'' in \emph{2021 IEEE/CVF Conference on Computer
  Vision and Pattern Recognition Workshops (CVPRW)}, 2021, pp. 691--700.

\bibitem{prabhakar21}
K.~R. Prabhakar, G.~Senthil, S.~Agrawal, R.~V. Babu, and R.~K. S.~S. Gorthi,
  ``Labeled from unlabeled: Exploiting unlabeled data for few-shot deep hdr
  deghosting,'' in \emph{2021 IEEE/CVF Conference on Computer Vision and
  Pattern Recognition (CVPR)}, 2021, pp. 4873--4883.

\bibitem{zaheer17}
M.~Zaheer, S.~Kottur, S.~Ravanbakhsh, B.~Poczos, R.~R. Salakhutdinov, and A.~J.
  Smola, ``Deep sets,'' in \emph{Advances in Neural Information Processing
  Systems}, vol.~30, 2017.

\bibitem{aittala18}
\BIBentryALTinterwordspacing
M.~Aittala and F.~Durand, ``Burst image deblurring using permutation invariant
  convolutional neural networks,'' in \emph{European Conference Computer
  Vision}, ser. Lecture Notes in Computer Science, vol. 11212.\hskip 1em plus
  0.5em minus 0.4em\relax Springer, 2018, pp. 748--764. [Online]. Available:
  \url{https://doi.org/10.1007/978-3-030-01237-3\_45}
\BIBentrySTDinterwordspacing

\bibitem{prabhakar19}
\BIBentryALTinterwordspacing
K.~R. Prabhakar, R.~Arora, A.~Swaminathan, K.~P. Singh, and R.~V. Babu, ``A
  fast, scalable, and reliable deghosting method for extreme exposure fusion,''
  in \emph{{IEEE} International Conference on Computational Photography, {ICCP}
  2019, Tokyo, Japan, May 15-17, 2019}.\hskip 1em plus 0.5em minus 0.4em\relax
  {IEEE}, 2019, pp. 1--8. [Online]. Available:
  \url{https://doi.org/10.1109/ICCPHOT.2019.8747329}
\BIBentrySTDinterwordspacing

\bibitem{sun18}
D.~{Sun}, X.~{Yang}, M.~{Liu}, and J.~{Kautz}, ``Pwc-net: Cnns for optical flow
  using pyramid, warping, and cost volume,'' in \emph{IEEE Conference on
  Computer Vision and Pattern Recognition}, 2018.

\bibitem{teed20}
Z.~Teed and J.~Deng, ``Raft: Recurrent all-pairs field transforms for optical
  flow,'' in \emph{European Conference on Computer Vision}, 2020.

\bibitem{kim19}
D.-W. Kim, J.~R. Chung, and S.-W. Jung, ``Grdn:grouped residual dense network
  for real image denoising and gan-based real-world noise modeling,'' in
  \emph{Computer Vision and Pattern Recognition Workshops}, 2019.

\bibitem{zhang18}
Y.~Zhang, Y.~Tian, Y.~Kong, B.~Zhong, and Y.~Fu, ``Residual dense network for
  image super-resolution,'' in \emph{Computer Vision and Pattern Recognition},
  2018.

\bibitem{johnson16}
J.~Johnson, A.~Alahi, and L.~Fei-Fei, ``Perceptual losses for real-time style
  transfer and super-resolution,'' in \emph{European Conference on Computer
  Vision}, 2016.

\bibitem{simonyan15}
K.~Simonyan and A.~Zisserman, ``Very deep convolutional networks for
  large-scale image recognition,'' in \emph{International Conference on
  Learning Representations}, 2015.

\bibitem{chen21}
J.~Chen, Z.~Yang, T.~N. Chan, H.~Li, J.~Hou, and L.-P. Chau, ``Attention-guided
  progressive neural texture fusion for high dynamic range image restoration,''
  2021.

\bibitem{tursun15}
O.~Tursun, A.~O. Aky{\"u}z, A.~Erdem, and E.~Erdem, ``An objective deghosting
  quality metric for hdr images,'' \emph{Computer Graphics Forum}, vol.~35,
  2016.

\bibitem{mantiuk21}
R.~Mantiuk and M.~Azimi, ``Pu21: A novel perceptually uniform encoding for
  adapting existing quality metrics for hdr,'' 06 2021, pp. 1--5.

\bibitem{wang04}
Z.~Wang, A.~Bovik, H.~Sheikh, and E.~Simoncelli, ``Image quality assessment:
  from error visibility to structural similarity,'' \emph{IEEE Transactions on
  Image Processing}, vol.~13, no.~4, pp. 600--612, 2004.

\bibitem{narwaria15}
\BIBentryALTinterwordspacing
M.~Narwaria, R.~K. Mantiuk, M.~P.~D. Silva, and P.~L. Callet, ``{HDR-VDP-2.2:}
  a calibrated method for objective quality prediction of high-dynamic range
  and standard images,'' \emph{J. Electronic Imaging}, vol.~24, no.~1, p.
  010501, 2015. [Online]. Available:
  \url{https://doi.org/10.1117/1.JEI.24.1.010501}
\BIBentrySTDinterwordspacing

\bibitem{pu20}
Z.~Pu, P.~Guo, M.~S. Asif, and Z.~Ma, ``Robust high dynamic range (hdr) imaging
  with complex motion and parallax,'' in \emph{Proceedings of the Asian
  Conference on Computer Vision (ACCV)}, November 2020.

\bibitem{yan20}
Q.~Yan, L.~Zhang, Y.~Liu, Y.~Zhu, J.~Sun, Q.~Shi, and Y.~Zhang, ``Deep hdr
  imaging via a non-local network,'' \emph{IEEE Transactions on Image
  Processing}, vol.~29, pp. 4308--4322, 2020.

\bibitem{grossberg04}
M.~Grossberg and S.~Nayar, ``Modeling the space of camera response functions,''
  \emph{IEEE Transactions on Pattern Analysis and Machine Intelligence},
  vol.~26, no.~10, pp. 1272--1282, 2004.

\bibitem{lee12}
J.-Y. Lee, Y.~Matsushita, B.~Shi, I.~S. Kweon, and K.~Ikeuchi, ``Radiometric
  calibration by rank minimization,'' \emph{IEEE Transactions on Pattern
  Analysis and Machine Intelligence}, vol.~35, no.~1, pp. 144--156, 2013.

\end{thebibliography}
\bibliographystyle{IEEEtran}

\end{document}